# Optical Phonons in Twisted Bilayer Graphene with Gate-Induced Asymmetric Doping

*Ting-Fung Chung,*[1,2] *Rui He,*[3] *Tai-Lung Wu,*[1,2] *Yong P. Chen*[1,2,4]

[1] Department of Physics and Astronomy, Purdue University, West Lafayette, IN 47907, USA

[2] Birck Nanotechnology Center, Purdue University, West Lafayette, IN 47907, USA

[3] Department of Physics, University of Northern Iowa, Cedar Falls, IA 50614, USA

[4] School of Electrical and Computer Engineering, Purdue University, West Lafayette, IN 47907, USA

ABSTRACT: Twisted bilayer graphene (tBLG) devices with ion gel gate dielectrics are studied using Raman spectroscopy in the twist angle regime where a resonantly enhanced G band can be observed. We observe prominent splitting and intensity quenching on the G Raman band when the carrier density is tuned away from charge neutrality. This G peak splitting is attributed to asymmetric charge doping in the two graphene layers, which reveals individual phonon self-energy renormalization of the two weakly-coupled layers of graphene. We estimate the effective interlayer capacitance at low doping density of tBLG using an interlayer screening model. The anomalous intensity quenching of both G peaks is ascribed to the suppression of resonant interband transitions between the two saddle points (van Hove singularities), that are displaced in the momentum space by gate-tuning. In addition, we observe a softening (hardening) of the R Raman band, a superlattice-induced phonon mode in tBLG, in electron (hole) doping. Our results demonstrate that gate modulation can be used to control the optoelectronic and vibrational properties in tBLG devices.

TEXT:

Recently there has been growing interest in two-dimensional (2D) van der Waals (vdW) materials and structures in which interlayer interaction can significantly affect these systems' properties and functionalities.[1-9] Twisted bilayer graphene (tBLG), in which the two graphene layers are stacked with a twist angle ($\theta$) and coupled by vdW force, has been demonstrated to show new physical (electronic, vibrational, and optical) properties through changed interlayer interaction at different twist angles.[10-15] Angle-resolved photoemission spectroscopy measurement has shown that tBLG exhibits weak interlayer coupling as revealed by the presence of van Hove singularities (VHSs) in the density of states at the overlap (saddle point) of two single layer graphene (SLG) Dirac cones.[13] Furthermore, low-energy, $\theta$-dependent VHSs and superlattice Dirac cones have been observed by scanning tunneling microscopy and spectroscopy.[16-18] Optical spectroscopy has been exploited to study the optical and vibrational properties associated with the low-energy VHSs of tBLG.[19-27] These studies have demonstrated that tBLG is a prototype system to explore the influence of interlayer interaction in 2D layered materials.

Raman spectroscopy is a sensitive probe of the unique electronic and phonon band structures of tBLG through resonance enhancement and superlattice induced Raman processes. Raman intensities from G and double-resonant (DR) ZO'$_L$ (fundamental layer-breathing vibration) bands display large resonance enhancements when the excitation photon energy equals to the inter-VHS energy (E$_{VHS}$; the energy difference between the saddle points in the conduction and valence bands).[19-22,27] The twist angle at which E$_{VHS}$ equals the excitation photon energy is called the critical angle $\theta_c$. For 532 nm laser excitation, $\theta_c$ is ~12.5°, and it becomes ~10.5° for 633 nm laser excitation.[22] Several new Raman bands, such as R, R' and ZO'$_H$, are activated by superlattice-induced wavevector.[25-27,29-30] These characteristic Raman features related to the low-energy VHSs and superlattice atomic structure have been observed only in tBLG,[19-31] but not in SLG or Bernal-stacked bilayer graphene (AB-BLG).

In this Letter, we report the observation of gate-induced G Raman band splitting and intensity quenching in tBLG with twist angle close to $\theta_c$, at which the ungated sample exhibits a large G band resonance enhancement. By creating an asymmetric doping in the two layers via electrochemical gating, the Raman spectra of tBLG evolve in ways that differ greatly from those observed for SLG and AB-BLG. The observed G band splitting shows no optical phonon mixing, suggesting the absence of the infrared (IR)-active E$_u$ mode that is present in gated AB-BLG devices.[32,33] In our studies, we are able to estimate the Fermi energy $E_F$ and carrier density in each graphene layer from the split G Raman bands which show individual phonon self-energy renormalizations (with different charge carrier densities in the two layers). An interlayer screening model is employed to explain the G band splitting of the tBLG with doping asymmetry. An effective interlayer capacitance of ~4.6 μFcm$^{-2}$ is estimated from the interlayer potential between two graphene layers. The unusual G Raman intensity quenching away from the CNP is attributed to the reduction in the joint density of states (JDOS) associated with interband transition near VHSs, in which the saddle points are displaced in both energy and momentum by the interlayer potential.[34] In addition, the dependence of the R Raman band on the gate voltage (carrier density) was observed for the first time. Its phonon self-energy renormalization could have contribution from both electron-phonon and electron-electron interactions, similar to what occurs in the 2D Raman band.

Our graphene samples were grown on Cu foils by chemical vapor deposition and transferred onto a heavily p-doped Si substrate (coated with ~300 nm $SiO_2$).[27] Field effect devices were fabricated for Raman studies and electrical characterizations. Figure 1(a) shows an optical image of the fabricated graphene device, which consists of SLG (upper) and tBLG (lower) regions (delimited by the dashed white line). A color contrast between tBLG (darker) and SLG can be seen. Raman studies from both the SLG and tBLG regions of the device were performed. From the R, G and 2D Raman characteristics, we determined that the twist angle $\theta$ of the tBLG is about 13°,[21-22,29] close to the critical angle $\theta_c$ (12.5° for the excitation photon energy of 2.33 eV of a 532 nm laser). The Raman G band intensity is strongly enhanced at this twist angle because the energy between the saddle points (VHSs) in the conduction and valence bands is resonant with the photon energy of the incident laser beam.[21,22] Figure 1(b) shows a schematic drawing of our device setup for electrochemical gating. A voltage ($V_{TG}$) applied to the side electrode is used to gate the graphene via the ion gel dielectrics (PEO:$LiClO_4$) that acts as a top gate. Details of the sample growth and experimental procedures can be found in the Suppl. Info.

Figure 1(c) displays three representative Raman spectra from tBLG at different $V_{TG}$. Spectra from the SLG are included for comparison. All spectra are normalized to the 520 $cm^{-1}$ Si peak. At $V_{TG}$ ~ 0.5 V, the R, G and 2D bands from the tBLG are at about 1492, 1584, and 2699

cm$^{-1}$, respectively. We estimate the charge neutral point (CNP) voltage $V_D \sim 0.5$ V based on the approximate symmetry of the spectra evolution with respect to electron (n-) and hole (p-) doping (also see Figs. 2 and 3a) away from this voltage. The non-zero $V_D$ is ascribed to unintentional extrinsic doping from the Si substrate and the ion gel electrolyte.[32,35] The positive and negative signs of the ($V_{TG} - V_D$) correspond to n- and p-doping in graphene, respectively. Doping dependence of the G and 2D bands from the SLG in our device is in good agreement with previous reports.[36-38] The $V_D$ of the SLG is determined to be $\sim 0.6$ V, slightly higher than that of the tBLG. This is also consistent with the $V_D$ values determined by electrical transport measurement (Fig. S5 in the Suppl. Info.). As shown in Fig. 1(c), the G Raman band from tBLG exhibits strong resonance enhancement (intensity $\sim 40$ times larger compared to that of SLG at the CNP).

Figures 1(c) and 2 show that the G band of the tBLG not only blue-shifts but also splits into two peaks when $V_{TG}$ is away from $V_D$. Near the CNP, the spectra are described by a single Lorentzian lineshape with a full-width-at-half-maximum (FWHM) of $\sim 15$ cm$^{-1}$, comparable to that of charge neutral SLG. In SLG the G band only shows a blueshift without splitting when $V_{TG}$ is tuned away from the CNP (see Fig. 1(c) and Fig. S1). This frequency upshift is well-studied and explained by phonon self-energy renormalization due to electron-phonon coupling (EPC).[36,38] Although a uniaxial strain may induce a splitting of the G band for the SLG,[39,40] such a G band splitting is not observed (see Figs. 1c and S1) in the SLG region in our device, indicating that strain is negligible in our fabricated devices. This further indicates that the observed G splitting in the tBLG (lower region in the same device, see Fig. 1a) is unlikely to be associated with strain.[41]

In gated AB-BLG device, it has been shown that the G band splits due to optical phonon mixing (symmetric $E_g$ and asymmetric $E_u$) when the AB sublattice symmetry is broken by application of an out-of-plane electric field, where the odd-parity $E_u$ mode becomes active in Raman scattering.[32,33] In tBLG, the AB sublattice symmetry is naturally broken because of the relative rotation between the two layers regardless of the charge doping. However, the $E_u$ mode has not been observed in Raman studies of tBLG under zero gate voltage,[21,22,32,35,42] suggesting this $E_u$ mode remains Raman-inactive or silent in tBLG. Araujo, *et al.* and Kalbac, *et al.* studied the Raman features of twisted bilayer $^{12}C/^{13}C$ graphene with large twist angle using electrochemical doping method.[35,42] No obvious signature of the $E_u$ mode has been observed. In addition, the doublet G lines observed in our doped tBLG sample are different from those reported on AB-BLG (due to optical phonon mixing) in which the two G Raman peaks give opposite frequency shift, while simultaneously a reversal of resonance intensities occurs with increasing doping density.[32,33] However, in our doped tBLG sample we observe a concurrent upshift of the doublet G lines (Figs. 1c and 2) and reduction of their intensities without crossing. Therefore, the G band splitting in our tBLG device is not caused by such optical phonon mixing. Instead, we attribute the splitting to the gate-induced asymmetric doping in the two layers of the tBLG, to be discussed in more details later.

Significant quenching of the resonantly-enhanced G Raman intensity with increasing doping level is also observed in the tBLG. Ratios of the integrated intensities of the G and 2D bands ($A_G/A_{2D}$) in the tBLG and SLG as functions of $V_{TG}$ are shown in the inset of Fig. 1(c). The $A_G/A_{2D}$ of the tBLG is at its maximum at ~ 1 V, which is significantly higher than the estimated $V_D$ of ~ 0.5 V, and then $A_G/A_{2D}$ drastically declines by a factor of up to 6 while the sample is heavily doped. In contrast, this $A_G/A_{2D}$ ratio in the SLG is at its minimum very close to the CNP ($V_D \sim 0.6$ V),[36] and it is enhanced by a factor of $\sim 3$ when the sample is heavily doped. The increase of $A_G/A_{2D}$ in SLG has been ascribed to the reduction of $A_{2D}$ (Fig. S2) due to an increase of scattering between photoexcited carriers as the doping level increases.[43,44] This doping dependence of the 2D intensity also occurs in the tBLG (Fig. 1c). However, the G band intensity may decrease even faster than the 2D intensity in tBLG as the doping level increases. Among all the Raman bands seen in the tBLG device (Fig. 1c), the G band intensity shows the strongest resonance at $\sim 1$ V (see Fig. 2b) $\sim 0.5$ V above the CNP), which may suggest that the energy separation between the VHSs is not well overlapped with the incident photon energy. This energy difference could be attributed to disorder (unintentional doping and strain caused by wrinkles) and the fact that the twist angle is slightly smaller or larger than the critical angle as $E_{VHS}$ is strongly dependent on resonant condition. The G band intensity is subject to the greatest suppression when the sample is further doped, implying strong influence of the doping on the resonance condition. Although the Fermi level in our experiment ($|E_F|$ can be tuned $\pm \sim 0.5$ eV away from CNP) cannot reach the $E_{VHS}$ ($\sim \pm 1$ eV from CNP) of the tBLG, the resonance condition can still be modulated by gating. The strong G Raman band intensity quenching in the tBLG is attributed to off-resonance or the reduced JDOS associated with the VHSs due to gating and will be discussed later in this paper.

Figures 3(a–c) show the peak frequencies $\omega_G$, FWHMs $\Gamma_G$, and integrated intensities $A_G$ of the doublet G peaks ($G_T$ and $G_B$, the subscripts T and B represent top and bottom layers, respectively) as a function of $V_{TG}$. It is reasonable to assign the layer which has more prominent changes in the G features as the top layer since this layer is in direct contact with the top ion gate electrolyte and is more strongly influenced by the gating. All parameters are extracted from simple fits of the bands with two Lorentzian peaks. A single Lorentzian function is used to fit the unsplit G peak in the vicinity of the CNP. These data points are shown by solid blue squares in Figs. 3(a–c). With increasing carrier density, the frequencies of the two G peaks blue-shift at different rates and their intensities decrease simultaneously, indicating the off-resonance condition when $V_{TG}$ is away from the CNP. These features are very different from the gate dependence of the G doublet peaks in AB-BLG in which the two G peaks repel each other in energy, and a reversal of their intensities takes place and crosses at around 200 meV with respect to the CNP. In our case of the doped tBLG, the two G peaks appear to be uncoupled to each other in frequency and FWHM, and show no crossing in their intensities. Unlike AB-BLG, the observed doping dependence of both $G_T$ and $G_B$ peaks in the tBLG agree quite well with those observed in SLG (Fig. S1) in which the frequency (FWHM) blueshifts (narrows) with increasing charge density. The $\Gamma_G$ changes by $\Delta\Gamma_G \sim 8.7 \pm 0.5$ cm$^{-1}$ for both the $G_T$ and $G_B$ peaks as $V_{TG}$ is tuned away from the CNP (Fig. 3b and inset of Fig. 3d). Following the similar method used to estimate the EPC strength in SLG from $\Delta\Gamma_G$,[38] we estimate that the EPC strength of each graphene layer in the tBLG is $14.3 \pm 0.4$ eV/Å, comparable to that of SLG ($\Delta\Gamma_G \sim 8.5$ cm$^{-1}$ and EPC strength of $\sim 14.1$ eV/Å).[38] This finding reveals that the interlayer interaction between the two graphene layers in the tBLG is sufficiently weak and has negligible effect on the EPC of the intralayer G phonons for each layer, which behaves similarly to a SLG.

We have calculated the carrier densities (doping) of each layer (top/bottom) from the corresponding G Raman peak ($G_T$/$G_B$) frequencies, assuming similar dependence of the G peak frequency as that found for a SLG. It has been experimentally shown that the G peak blue-shifts linearly with $E_F$ in SLG ($\omega_G \propto E_F$).[38,45] This feature is confirmed in our SLG (Fig. S1, Eqs. S3 and S4 in Suppl. Info.), yielding a linear relation $|E_F| \times 40 = \omega_G - 1583.8$ ($|E_F| \times 45 = \omega_G - 1583.8$) for n- (p-) doped SLG, where $E_F$ and $\omega_G$ are in units of eV and cm$^{-1}$, respectively, in good agreement with prior studies.[45] Taking into account of electron-hole asymmetry and different minimum G Raman peak frequencies (~2 cm$^{-1}$) in tBLG (~1583 cm$^{-1}$) and SLG (~1585 cm$^{-1}$), we extract the $E_F$ and carrier concentration $n = (E_F/\hbar v_F)^2/\pi$ (linear energy dispersion) in the top ($n_T$) and bottom ($n_B$) layers of the tBLG (Fig. 3d) using modified relations in the form of $|E_F(n_T)| \times 42 = \omega_G(n_T) - 1582$ and $|E_F(n_B)| \times 42 = \omega_G(n_B) - 1582$, respectively. The error bars (Fig. 3d) include the uncertainties of the two numerical values used (42 and 1582). The total carrier concentration $n_{total} = n_T + n_B$ of the tBLG system is shown as empty black circles in Fig. 3d. The charge density in the SLG (denoted as $n_{SLG}$) is also calculated from its G Raman frequency (Eqs. S3 and S4) and shown as empty blue triangles. Finally, the effective charge density (denoted as $n_{TG}$) induced by the ion gel gating on SLG is calculated by $e(V_{TG} - V_D) = n_{TG}e^2/C_{TG} + \hbar v_F\sqrt{n_{TG}\pi}$ (see Eq. S2 in Suppl. Info.) and shown as a solid gray line in Fig. 3d,[36] where $C_{TG} \approx 2$ μFcm$^{-2}$ is the capacitance of the electrolyte and agrees with prior reports,[36] $V_{TG} - V_D$ is the applied voltage relative to CNP, $e$ is electron charge. The first and second terms are ascribed to geometric and quantum capacitances, respectively. At low doping (ΔV ~ ± 2 V; equivalent to $n$ ~ 1.8 × 10$^{13}$ cm$^{-2}$), $n_{total} = n_T + n_B$ for the tBLG calculated from the $G_T/G_B$ Raman peaks agrees well with $n_{TG}$ and $n_{SLG}$. When |ΔV| > ±2 V, $n_{total} = n_T + n_B$ deviates notably from both $n_{SLG}$ and $n_{TG}$, possibly due to a reduced gating efficiency of electrolyte at relatively high gate voltages. We also notice more electron-hole asymmetry in $n_{total}$ at such large gate biases. The consistency between the doping density extracted from G Raman peaks (based on the assumption that each layer behaves as SLG) of the tBLG and those expected from the capacitance and measured from the SLG confirms that the coupling between the two layers in our tBLG system is sufficiently weak such that each layer retains its SLG-like low energy electronic structure (Dirac band dispersion) and phonon self-energy renormalization (dependence of G peak frequency on $E_F$). On the other hand, we point out that the coupling between the two layers still exists, giving rise to the VHSs at higher energies due to the coupling between Dirac cones from the two graphene layers, as manifested by the resonantly enhanced G band.

The difference in the gate-dependence of the $G_T$ and $G_B$ peaks also reflects the difference in the phonon renormalization magnitudes due to different carrier densities in the two graphene layers. As shown in Fig. 3(d), the carrier density in the bottom layer ($n_B$) becomes almost constant around $\pm 0.4 \times 10^{13}$ cm$^{-2}$ when $|V| > 2$ V, and additional doping mainly contributes to the top layer. This leads to continued increase in the peak frequency of $G_T$ but saturation of the peak frequency of $G_B$ upon further increasing of $|V|$. The inset of Fig. 3(d) displays the $\Gamma_G$ of the doublet G peaks as a function of the Fermi energy ($E_{F\_\omega_G}$ estimated from the $G_T$ and $G_B$ phonon frequencies) in each layer. We note the similarity of the lineshape between the top and bottom layers within $E_{F\_\omega_G} = \pm 0.2$ eV. Furthermore, the widths of the two $\Gamma_G$ vs. $E_{F\_\omega_G}$ peaks are close to the phonon energy $\hbar\omega_G$ (~ 200 meV), indicating Landau damping of the G phonons which decay into electron-hole pairs.[38]

Figures 3(e–g) schematically illustrate an interlayer screening model that we propose to describe the G band splitting and Raman intensity quenching observed in tBLG. The charge distribution over the top ($n_T$) and bottom ($n_B$) layers depends on the electrostatic interaction between layers and band-filling.[46,47] Both the top and bottom graphene layers are in direct contact with each other and with the metal electrodes. Therefore, the $E_F$'s of the two layers are assumed to be aligned when the system is in equilibrium. In the undoped tBLG (ideal flat band condition), there is no potential difference ($\Delta\phi$) between the two layers, and the $E_F$ is at the CNP.

In this case, the conduction and valence bands near the saddle points (VHSs) of tBLG are aligned and parallel to each other, maximizing the JDOS for resonant interband transitions (green arrows) between the VHSs,[10,22,34] and a very strong enhancement of the G Raman band appears.

An accumulation of positive ions in the electrolyte results in n-doped tBLG (Fig. 3f). The doping is more efficient in the top layer because the electrolyte ions are closer to the top layer than to the bottom layer (in contact with the Si substrate). The two layers of the tBLG share the same aligned $E_F$ (dashed red line). However, their CNPs are lifted by an interlayer potential difference ($\Delta\phi$). The top and bottom layers of the tBLG feel different electric fields $E_T = (n_T + n_B)e/\varepsilon_{PE}\varepsilon_0$ and $E_B = n_B e/\varepsilon_G \varepsilon_0$, where $\varepsilon_{PE}$ and $\varepsilon_G$ are the relative dielectric constants of the electrolyte and graphene, respectively, $\varepsilon_0$ is the vacuum permittivity. The difference in the electric field ($E_T - E_B > 0$) is attributed to electronic screening by the charge carriers of the top layer. Indeed, the electronic screening plays a crucial role in creating charge density asymmetry in graphene layers, and the strength of the screening depends on the doping level as studied by Kuroda et al.[47] The screening length corresponding to our doping level of $10^{13}$ cm$^{-2}$ is only a fraction of the graphene interlayer spacing ($d_0 \approx 0.34$ nm).[47] The strong resonance enhancement on the G Raman band in the flat band case originates from the resonant interband transitions between the saddle points in the absence of the interlayer potential.[21,22] In the presence of the interlayer potential, the saddle points are oppositely displaced in momentum space and the electronic band structure is altered (Fig. 3f).[10,34,48] Direct interband transitions connecting the two saddle points (VHSs) become forbidden (in this sense the energy separation between the saddle points become "indirect", as demonstrated by the dashed green arrows, in analogy with an indirect bandgap in semiconductors). Therefore, the JDOS of the system for the interband transition (between VHSs) and the resonant G band enhancement are suppressed. Note that this mechanism is different from the modification of JDOS (optical absorption) caused by the many-body effects (electron-hole and electron-electron interactions) in doped SLG.[49] Similar explanation is applicable to the p-doped tBLG (Fig. 3g).

We can quantitatively describe the $n_T$ and $n_B$ dependence on $V_{TG}$ using the band diagrams shown in Figs. 3(f–g). An applied $V_{TG}$ is the sum of potential drop across the Debye length of the electrolyte[36] (due to electrostatic capacitance) and the Fermi energy (with respect to CNP) of the top layer (due to quantum capacitance): $eV_{TG} = e^2(n_T + n_B)/C_{TG} + E_F(n_T)$, here $E_F(n_T)$ is positive (negative) for electron (hole) carriers. Similarly, the Fermi energy of the top layer can be written as the sum of the Fermi energy (with respect to CNP) of the bottom layer and the interlayer potential: $E_F(n_T) = E_F(n_B) + \phi$. If we treat the two layers of tBLG as a simple parallel-plate capacitor, the interlayer potential is $\phi = E_F(n_T) - E_F(n_B) = e^2 n_B/C_{tBLG}$ (Fig. S3), where $C_{tBLG}$ is the effective interlayer dielectric capacitance per unit area of graphene, originating from the effect of electronic screening ($V_{TG}$ 0). Note that $\phi$ is positive (negative) in n- (p-) doped tBLG. From the above analysis, we determine the effective interlayer static capacitance $C_{tBLG} \sim 4.6$ μFcm$^{-2}$ from the slopes (linear blue lines for both carriers) in Fig. S3, close to $5.2 - 7.8$ μFcm$^{-2}$ estimated from $C_{tBLG} = \varepsilon_G \varepsilon_0/d_0$. Here, the relative dielectric constant of BLG is $\varepsilon_G = 2 - 3$,[50] and $d_0 = 0.34$ nm is used in this estimation.

We note that there are other anomalous features in our data. First, the CNP of the top and bottom layers in the tBLG are slightly different (by $\sim 0.2$ V) (see Fig. 3a). This asymmetry may be attributed to unintentional doping by the substrate and non-uniform doping by the polymer electrolyte. The bottom layer is in direct contact with thus subject to a stronger influence from the substrate. It has been shown that charged impurities can be trapped at the tBLG/substrate interface in the graphene transfer process. These impurities may cause the two graphene layers to respond differently during gating.[51] In addition, the polymer electrolyte may dope the two graphene layers differently at $V_{TG} = 0$ (the top layer is doped with more carriers on the order of ~ $10^{12}$ cm$^{-2}$ since it is in contact with the electrolyte).[35] Second, there are discontinuities in the gate dependent G frequencies, FWHMs, and integrated intensities when $V_{TG}$ ~ -1.5 V ($E_F$ ~ −0.4 eV) (see Figs. 3a–d). Prior experiments on AB-BLG showed a kink in the G Raman frequency at $E_F$ ~ 0.4 eV which is associated with second sub-band filling.[37] Theoretical studies suggest an absence of sub-band between the VHSs of tBLG.[11,12,48] Further studies are needed to understand the origin of these kinks in tBLG.

We now discuss the influence of asymmetric doping on the 2D and R Raman bands from the tBLG. These two bands are activated by intervalley DR process with phonon wavevector $q \neq 0$.[25,29,52] The 2D and R bands come from the same TO phonon branch but at different locations of the Brillouin zone (BZ). The 2D band originates from the scattering between the two adjacent Dirac cones (K and K') of a graphene layer with phonon wavevector $q$ which is equal to the K-K' separation (same as Γ- K separation in the BZ). The R band has a smaller phonon wavevector which equals the tBLG superlattice wavevector (see inset in Fig. 4b).[25,26]

Figure 4(a) shows the evolution of phonon frequency and FWHM of the 2D band as a function of $V_{TG}$. All parameters are extracted from simple fits of the 2D band with single Lorentzian peaks. We obtain $\partial\omega_{2D}/\partial E_F$ ~ − 7 cm$^{-1}$eV$^{-1}$ for n-doping and weak doping dependence for p-doping in the tBLG. However, in the SLG (Fig. S2) we obtain $\partial\omega_{2D}/\partial E_F$ ~ 23 cm$^{-1}$eV$^{-1}$ for both n- and p-doping from our measurement, which is consistent with previous reports.[36] The doping dependence of the FWHM of the 2D band in tBLG and SLG are different in p-doping, but similar in n-doping. In high p-doping regime, the FWHM of the tBLG increases by ~10 cm$^{-1}$ compared to its value at the CNP, but there is only a small variation (less than 3 cm$^{-1}$) in the FWHM of the 2D band in the p-doped SLG (see Fig. S2). The difference between the doping dependence of the 2D band of the tBLG and SLG may be linked to the difference in their band structures under an electric field. A change in the band structure of tBLG due to interlayer potential may have a large impact on the interaction of this second-order phonon with photons and electrons. Further studies are required to understand the mechanism causing the differences in the 2D band between tBLG and SLG.

A recent study of tBLG devices in which the twist angles are slightly smaller than the critical angle and electrodes are only in contact with one of the layers found that the 2D Raman band displays an asymmetric lineshape that can be decomposed into two peaks with similar widths.[53] The 2D splitting is attributed to different scattering pathways in double-resonance process near the saddle points in the electronic band structure of tBLG.[53] We did not observe such 2D band splitting in our tBLG devices. Precise reasons for this difference remain to be better understood but it may be related to several factors. First, the twist angle (~ 13°) of our tBLG is slightly larger than the critical angle (~ 12.5°) measured with 532 nm excitation laser energy. Second, both the top and bottom layers of our tBLG are in contact with the electrodes, which facilitate the alignment of $E_F$ of the two layers when a $V_{TG}$ is applied. In addition, the ion gel dielectrics (PEO:LiClO$_4$) we used gives higher carrier densities (~ 3 × 10$^{13}$ cm$^{-2}$) compared to those with Si backgate (~ 1 × 10$^{13}$ cm$^{-2}$).[35,36,53] These differences may lead to distinct Raman features in our tBLG compared to those reported in recent literature.[35,42,53]

Figure 4(b) plots the frequency and FWHM of the R band as a function of $V_{TG}$. The R band is too weak to be detected for heavily p-doped regime ($V_{TG}$ < -1.5 V, Fig. 1c) and can only be fitted with a single Lorentzian function. The observed gate dependence of the R band frequency is similar to that of the 2D band. This may be linked to the fact that both bands are from the same TO phonon branch. We obtain $\partial\omega_R/\partial E_F \sim -9.2$ cm$^{-1}$eV$^{-1}$ for n-doping which is slightly larger than that obtained from the 2D band, and $\partial\omega_R/\partial E_F \sim 3.8$ cm$^{-1}$eV$^{-1}$ for p-doping. The similar gate dependence of the R and 2D peak frequencies suggests that the phonon self-energy renormalization for the R band could share similar scattering mechanisms (e.g. a combination of electron-phonon and electron-electron interactions) as the 2D band.[44] The gate-dependent frequency shift can be expressed using a phenomenological formula based on DFT calculation[54]: $\omega = a + bn_{TG} + cn_{TG}^2 + dn_{TG}^3 + e|n_{TG}|^{3/2}$, where $\omega$ is the phonon frequency, $n_{TG}$ (~ $n_{total}$ in low doping regime) is the effective carrier density (in the units of $10^{-13}$ cm$^{-2}$) and $a$, $b$, $c$, $d$, and $e$ are coefficients. Fittings of gate dependence of the R and 2D frequencies to this phenomenological formula are shown in Fig. S4, and the fitting parameters are summarized in Table S1 in the Suppl. Info. Prior studies have proposed that the twist angle of tBLG can be estimated from the frequencies of Raman R and R' bands.[26,29,55] Our result on the R band suggests that doping level should be taken into account when determining the twist angle of tBLG via Raman measurements.

Although the gate dependence of the frequencies of the 2D and R bands are similar, the dependence of the FWHM of the two bands on the doping level is very different. The FWHM of the R band ($\Gamma_R$) reaches a maximum of ~8 cm$^{-1}$ at $V_{TG}$ ~ 1 V (~ 0.5 V away from the CNP) and then decreases for $V_{TG}$ away from this value, including both highly p- and n-doped regimes (Fig.

4b). In contrast, the FWHM of the 2D band from tBLG shows a minimum at ~ 0 V and increases rapidly in the p-doped regime. Further work is required to understand this difference.

In summary, novel features of the G Raman band were observed in tBLG under gate tuning. In the presence of doping asymmetry (interlayer potential) in the two layers, a splitting of G Raman peak was observed. We determined the $E_F$ and carrier concentration in each layer from the positions of the two G peaks. We also observed a strong gate-dependent quenching of the G peak intensities. It is interpreted by the suppression of interband direct transitions associated with the two low-energy saddle points (VHSs), which are oppositely shifted by interlayer potential, in the electronic structure of tBLG. An interlayer screening model was proposed to describe the observed phenomena, giving the effective interlayer capacitance of ~ 4.6 µFcm$^{-2}$. The similarity of the gate dependence of the 2D and R frequencies suggests that the phonon self-renormalization of the R and 2D bands could share similar scattering mechanisms. Our findings demonstrated that doping asymmetry significantly alters the properties of tBLG. This gate modulation can therefore be used to control the physical properties of tBLG devices.

Supporting Information: The material contains details about experimental procedures (graphene sample preparation, electrochemical top-gate field effect measurement and Raman spectroscopy), the method in determining the linear dependence of the $E_F$ on the G peak frequency in SLG, the doping dependence of the G and 2D bands in the SLG, the interlayer capacitance fitting from the interlayer potential in the bilayers, the fitting and corresponding parameters for the carrier density dependence of the 2D and R Raman peak frequencies in the tBLG, and the electrical field effect measurement of the SLG and tBLG devices.

ACKNOWLEDGMENTS: Y. P. C. acknowledges support from DTRA. R. H. acknowledges support by the American Chemical Society Petroleum Research Fund (Grant 53401-UNI10), NSF grant (DMR-1410496), and the UNI Faculty Summer Fellowship.

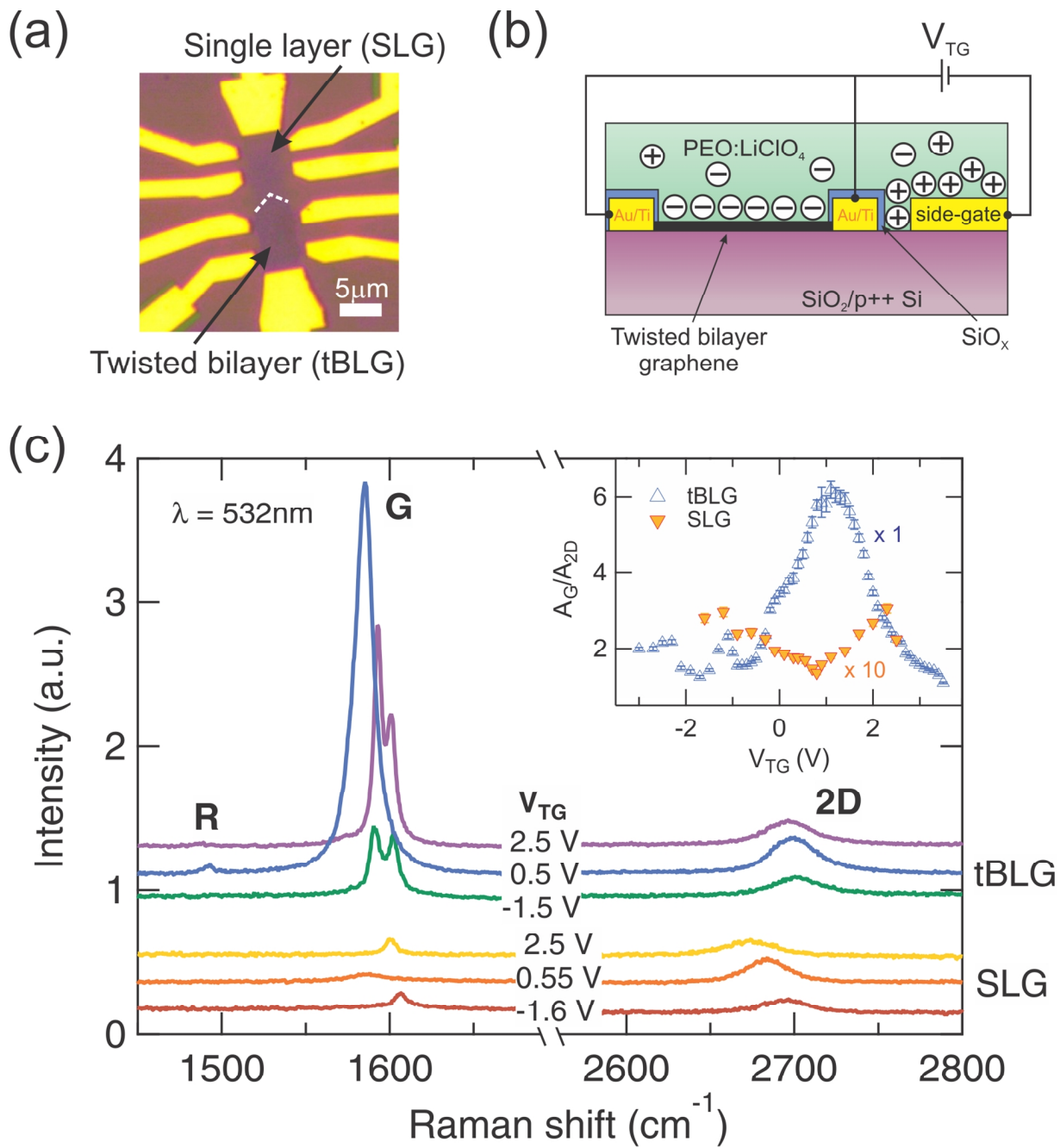

**Figure 1.** (a) An optical image of an electrochemically top-gated single layer (SLG) and twisted bilayer graphene (tBLG) device before application of ion gel electrolyte. The sample consists of a SLG (upper part) and a tBLG (lower part). The two parts show different optical contrast. The boundary between the SLG and tBLG is delimited by a dashed white line. (b) Schematic of device configuration (for the case with negative electrolyte top-gate voltage $V_{TG}$). (c) Comparison of Raman spectra of the SLG and tBLG at several different gate voltages $V_{TG}$. Spectra are normalized to the height intensity of the 520 cm$^{-1}$ silicon peak and are shifted vertically for clarity. All data were taken at room temperature using a 532 nm laser excitation. The charge neutrality point (CNP) voltage ($V_D$) of the SLG and tBLG is ~ 0.6 V and ~ 0.5 V, respectively, as estimated from the minimum of G band frequency (Fig. 3a). The sample is electron (n)-doped for $V_{TG} > V_D$, and is hole (p)-doped otherwise. The vertical scale is the same before and after the break on the horizontal axis. The upper right inset shows the ratios of the integrated intensities of the G and 2D peaks ($A_G/A_{2D}$) as a function of $V_{TG}$ from both the SLG and tBLG. The data of the SLG in the inset is multiplied by a factor of 10 for clarity.

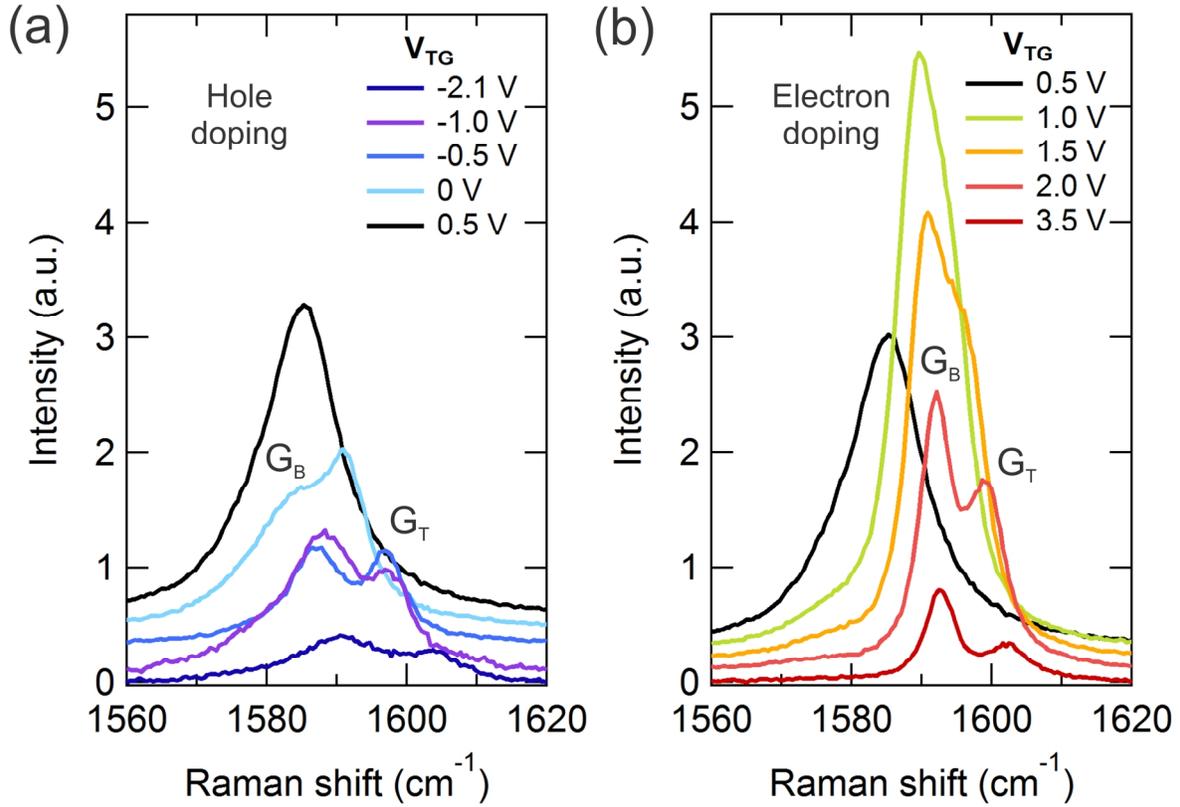

**Figure 2.** (a) Evolution of the normalized Raman spectrum in the region of the G band in the tBLG as a function of $V_{TG}$ in p-doped regime. (b) Same as in (a) for n-doped regime. The spectra at $V_D \sim 0.5$ V are plotted with black line. The doublet G bands are denoted as $G_B$ and $G_T$ peaks in which the subscripts B and T represent the bottom and top graphene layer, respectively, in the tBLG. Spectra are shifted vertically for clarity.

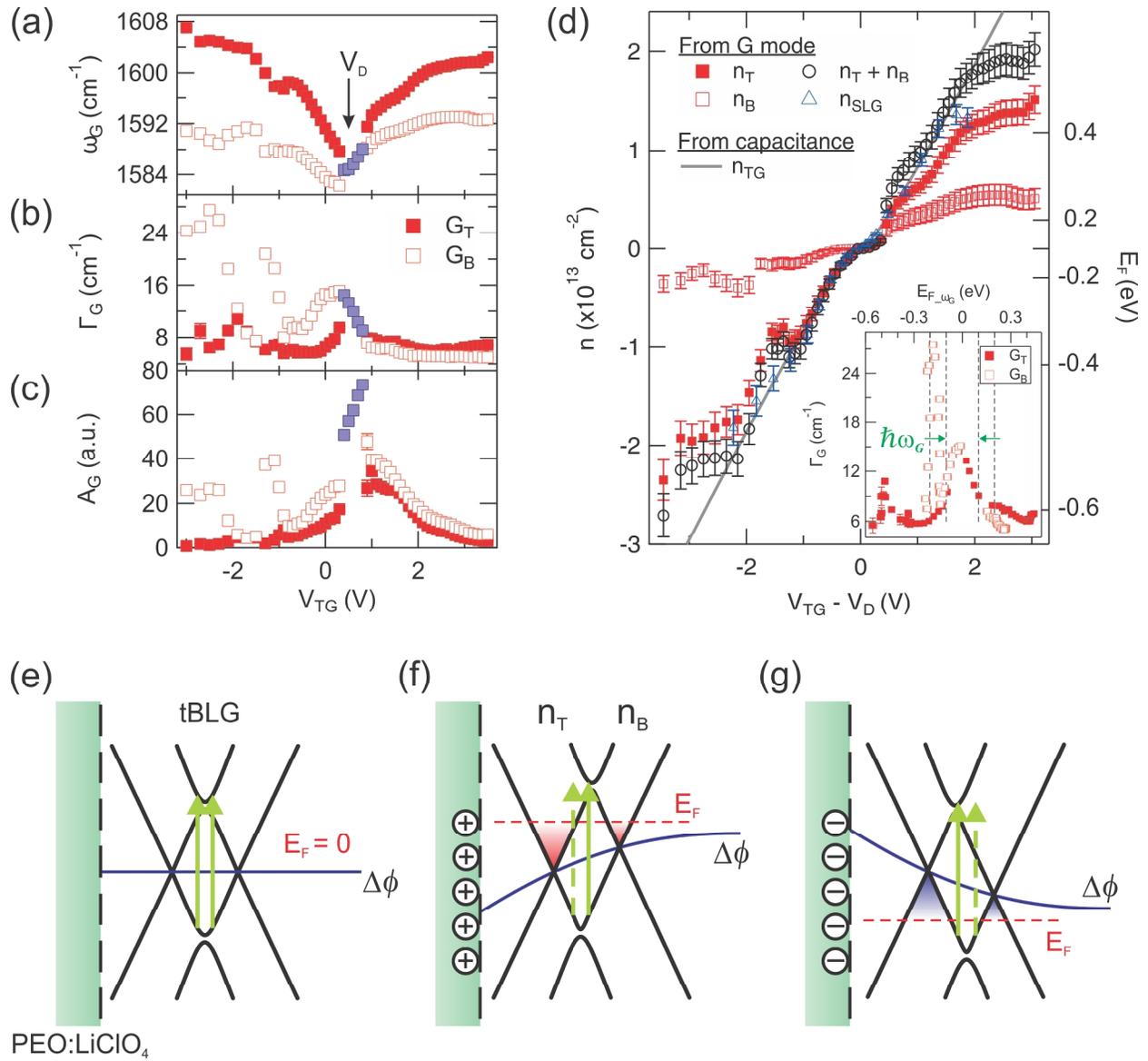

**Figure 3.** (a – c) Evolution of the frequency ($\omega_G$), FWHM ($\Gamma_G$), and integrated intensity ($A_G$) of the $G_B$ and $G_T$ peaks as a function of $V_{TG}$. The blue solid squares correspond to G peaks that show single Lorentzian lineshape (no splitting). (d) Carrier densities (doping) calculated either from Raman G peak position assuming SLG behavior, or from gate capacitance. The total density of the tBLG ($n_T + n_B$) is in good agreement with that of SLG ($n_{SLG}$). The induced carrier density in SLG estimated from the gate and quantum capacitances ($n_{TG}$, Eq. S2) is plotted for comparison. The inset shows the evolution of the $G_B$ and $G_T$ FWHMs as a function of the effective Fermi energy $E_{F\_\omega_G}$ of each individual layer. (e) Schematic energy band diagram of tBLG when it is undoped. Significant interband transitions (solid green arrows) are indicated which give rise to strong resonance enhancement on the G band. (f) Same as in (e) for n-doped situation assuming the two layers are in equilibrium (same chemical potential indicated by $E_F$). Electric-field screening results in an interlayer potential offset ($\Delta\phi$) between the layers, resulting in the higher charge carrier density in the top layer ($|n_T|>|n_B|$). The dashed green arrow shows the interband direct transition which becomes forbidden due to the shift of the two Dirac cones, leading to the intensity quenching of the G bands. (g) Same as in (f) for p-doped situation.

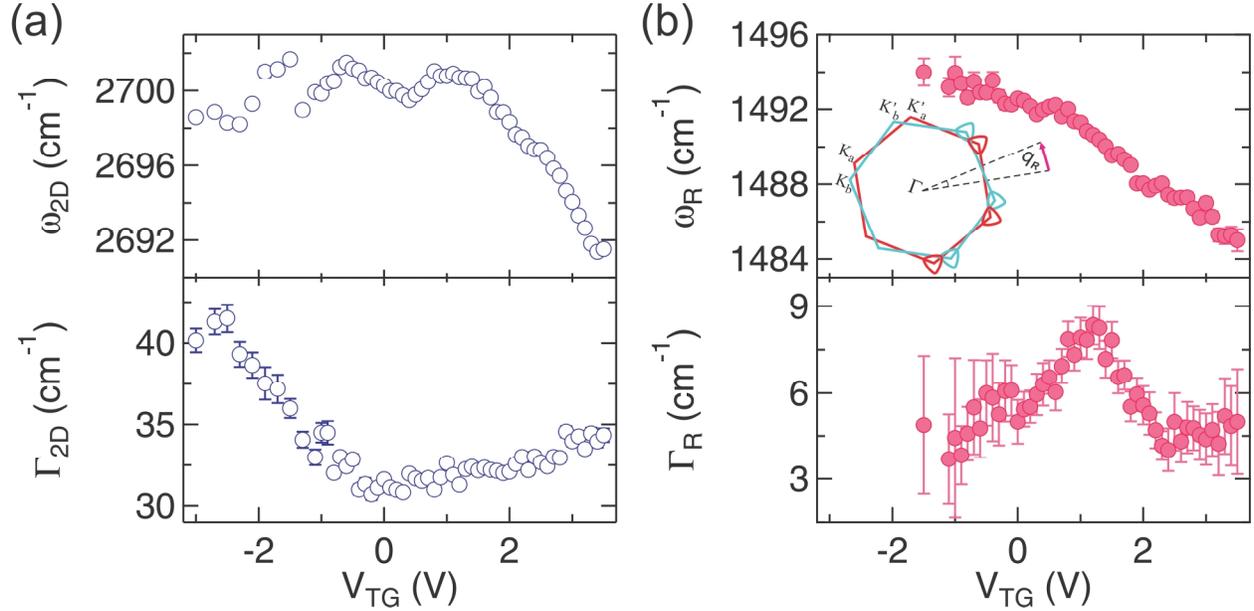

**Figure 4.** (a) and (b) Peak position ($\omega_{2D}$, $\omega_R$) and FWHM ($\Gamma_{2D}$, $\Gamma_R$) of 2D (left panel) and R (right panel) Raman bands as a function of $V_{TG}$. The inset in (b) shows the first Brillouin zones of the top and bottom graphene layers rotated from each other by a twist angle of ~13°. Wavevector of R phonon ($q_R$) is labeled.

# Supplementary Information

## 1. Experimental procedures

### Graphene sample preparation

We grew twisted bilayer graphene (tBLG) islands on copper foil using chemical vapor deposition (CVD) method and transferred the graphene layers onto a heavily doped silicon substrate with ~ 300 nm thermal oxide.[S1-S3] We identified bilayer graphene islands based on their color contrast under optical microscope.[S4] Then we used Raman spectroscopy (Horiba Xplora Raman spectrometer) to confirm that the bilayers have strong G band enhancement and sharp R peak (~ 1485 – 1500 $cm^{-1}$) using a 532 nm laser excitation source. E-beam lithography was utilized to define device pattern and electrodes. Excess graphene areas were removed by oxygen plasma etching. We evaporated 1/50 nm Ti/Au for contact and side-gate electrodes. We also evaporated 30 nm $SiO_2$ to cover electrodes in order to eliminate the contact between electrodes and ion gel electrolyte (for electrochemical gating). Device configuration is shown in Fig. 1(b) in the main text.

### Electrochemically top-gated field effect measurement

We employed the ion gel electrolyte (PEO:LiClO$_4$, also known as solid polymer electrolyte) as top-gate dielectrics which can electrically induce carrier density as high as ~ $4 \times 10^{13}$ $cm^{-2}$ in our experiment. The electrolyte solution used was PEO:LiClO$_4$ = 8:1 in weight, as in previous studies.[S5] The estimated gate capacitance $C_{TG}$ of the electrolyte is ~ 2 $\mu Fcm^{-2}$.[S5] A small drop of polymer electrolyte was applied on the graphene devices after all the electrodes were fabricated. Electrochemical doping of the graphene device was carried out by varying the applied gate voltage ($V_{TG}$) through the side gate electrode embedded in the electrolyte layer.

Field effect measurements were carried out before and after application of the electrolyte using a low-frequency lock-in amplifier (SRC-830) and a Keithley 2400 source meter. The measurements were conducted in vacuum (~ $10^{-4}$ Torr) at room temperature. Typical field effect curves of the tBLG and single layer graphene (SLG) are presented in Fig. S5, showing ambipolar characteristics with field effect mobility $\mu_{FE}$ of 6000 – 6500 cm$^2$V$^{-1}$s$^{-1}$, and 2800 – 3100 cm$^2$V$^{-1}$s$^{-1}$ at carrier density ~ 3 × 10$^{12}$ cm$^{-2}$ (from Si backgate) for the tBLG and SLG, respectively.

**Raman spectroscopy with an electrochemical top-gate**

Raman spectra at various $V_{TG}$ were measured using a LabRam HR spectrometer (Horiba) with 532 nm (2.33 eV) laser source and power level of ~ 2.5 mW. A long working distance objective lens (Olympus SLM-100×, N.A. 0.6) and an 1800 lines/mm grating were used in the gate-dependent Raman studies. The spectral resolution of the setup is about 0.5 cm$^{-1}$. The polymer electrolyte gating was achieved using a Keithley 2400 source meter. In our experiments, the $V_{TG}$ was changed from positive to negative. The sign of positive and negative $V_{TG}$ corresponds to electron (n-) and hole (p-) doping, respectively. At each $V_{TG}$ step, a Raman spectrum was taken while keeping the $V_{TG}$ constant. All the measurements were performed in air at room temperature.

## 2. Determination of graphene Fermi energy using Raman spectroscopy

Figures S1(a–c) plot the Raman shift, full-width-at-half-maximum (FWHM) of the G peak and the ratio of the integrated intensities of 2D and G peaks as a function of the carrier density $n$ in the SLG device. We convert $V_{TG}$ into effective doping density $n_{TG}$ (in consideration of quantum capacitance) by the following relation.[S5] The application of $V_{TG}$ generates potential differences through geometric and quantum capacitances.

$$e(V_{TG} - V_D) = n_{TG}e^2/C_{TG} + \hbar v_F\sqrt{n_{TG}\pi} \qquad (S1)$$

where $V_D$ is the CNP voltage of graphene, $v_F$ ~ 1 × 10$^8$ cms$^{-1}$ is the Fermi velocity, $C_{TG}$ ~ 2 × 10$^{-6}$ Fcm$^{-2}$ is the estimated geometric gate capacitance per unit area of the ion gel electrolyte. Using the numerical values, the equation becomes

$$e\Delta V_{TG} = 1.167 \times 10^{-7}\sqrt{n_{TG}} + 8.011 \times 10^{-14} n_{TG} \qquad (S2)$$

This equation allows us to estimate the $n_{TG}$ (in units of cm$^{-2}$) at each $V_{TG}$, as shown in Figs. S1(a–c).

We consider our tBLG (twist angle of ~ 13°) as two weakly-coupled graphene layers. The Fermi energy ($E_F$) of the tBLG with two split G Raman bands is estimated using the fact that the G peak frequency ($\omega_G$) of SLG is linearly dependent on its $E_F$.[S6,S7] This linear relationship was first determined from the Raman results measured on the SLG (see Fig. 1a in the main text for an image of the SLG that is used as a reference in this work. It is connected to the tBLG that we studied). Figure S1(d) shows the $\omega_G$ as a function of its $E_F = \hbar v_F \sqrt{n\pi}$ (linear energy dispersion) of the SLG. At high doping ($E_F \gg \hbar \omega_G$), $E_F$ is linearly proportional to the change of $\omega_G$, $A|E_F| = \omega_G$, where A is a constant. Therefore we obtain the relation between $E_F$ and $\omega_G$ for SLG, which is consistent with previous report:[S7]

$$|E_F| = \frac{\omega_G - 1583.8}{45} \text{ (eV) for p-doped SLG} \quad (S3)$$

$$|E_F| = \frac{\omega_G - 1583.8}{40} \text{ (eV) for n-doped SLG} \quad (S4)$$

where $E_F$ is in units of eV and $\omega_G$ is in units of cm$^{-1}$. These equations and the linear energy dispersion ($E_F = \hbar v_F \sqrt{\pi n}$) of SLG enable us to estimate the $E_F$ and corresponding $n$ at each $V_{TG}$ from $\omega_G$ in this study.

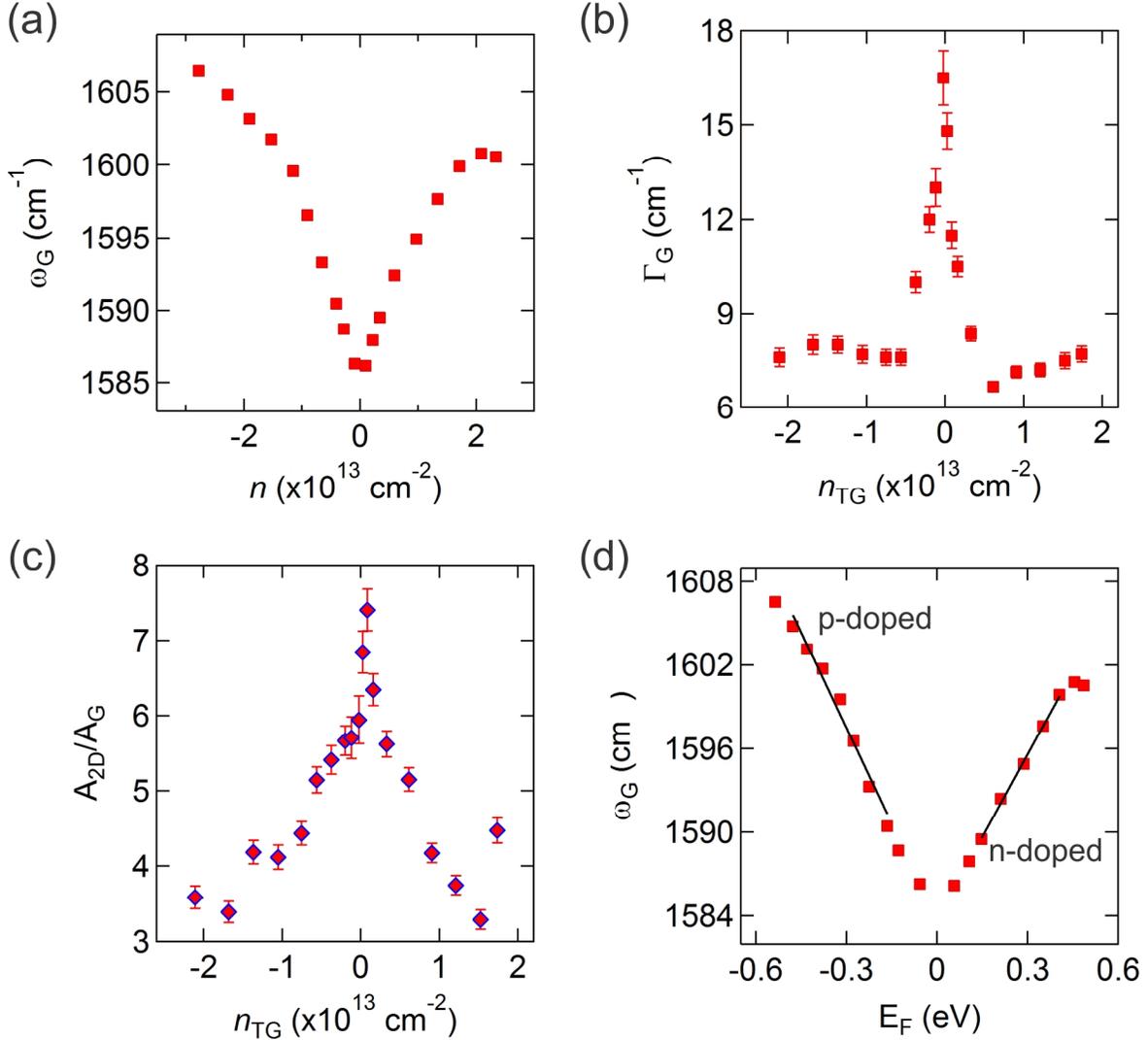

Figures S1. Dependence of the G band frequency (a), FWHM (b), and integrated intensity ratio $A_{2D}/A_G$ (c) on the carrier density $n_{TG}$ (calculated from the gate voltage based on geometric and quantum capacitances, Eq. S2), measured in the SLG device shown in Fig. 1 in the main text. The positive and negative signs of $n_{TG}$ denote electron (n-) and hole (p-) doping in SLG, respectively. All the features are consistent with typical carrier density dependence of SLG.[S5,S6] (d) Linear dependence of $\omega_G$ on $E_F$ in the p-doping and n-doping regimes of the SLG device. Experimental data (solid red squares) are fit with straight (black) lines. We estimate that $\partial\omega_G/\partial E_F \approx 45\,\mathrm{cm^{-1}eV^{-1}}$ for p-doping and $\partial\omega_G/\partial E_F \approx 39\,\mathrm{cm^{-1}eV^{-1}}$ for n-doping. The doping efficiency is comparable to that in Ref. S5.

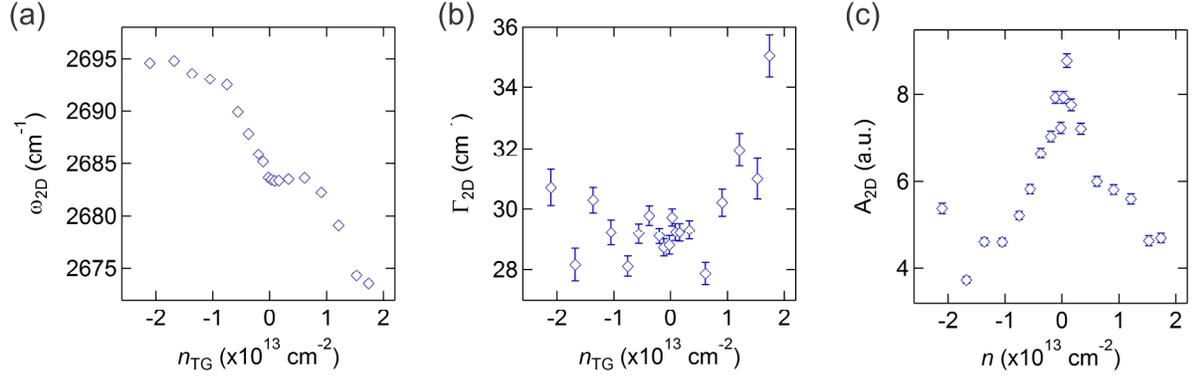

Figures S2. Dependence of the 2D band frequency (a), FWHM (b), and integrated intensity (c) on $n_{TG}$ in the SLG sample shown in Fig. 1 in the main text. The spectra are normalized to the Si peak at 520 cm$^{-1}$.

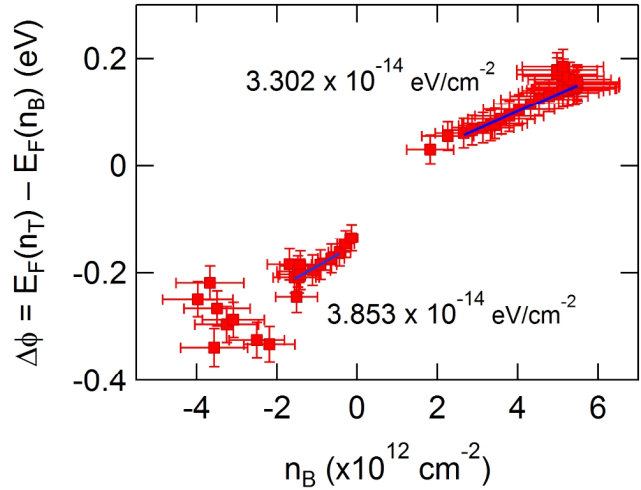

Figures S3. Interlayer potential $\phi = E_F(n_T) - E_F(n_B)$ between the top and bottom graphene layers versus the carrier density of the bottom layer ($n_B$) in the tBLG device. The Fermi energies measured from the charge neutrality point (CNP) of the top and bottom layers are denoted as $E_F(n_T)$ and $E_F(n_B)$, respectively. The effective interlayer static capacitance ($1/C_{tBLG} = \phi/e^2 n_B$) of the gated tBLG is obtained from the linear fits (solid blue lines) of the data points which are away from the CNP (> $10^{11}$ cm$^{-2}$).

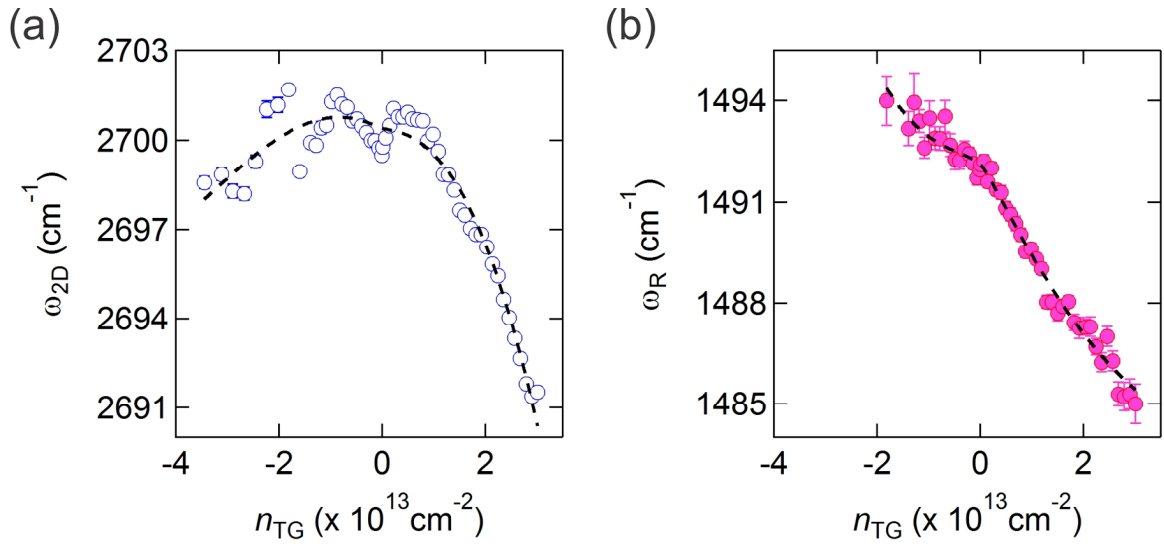

Figures S4. Peak position ($\omega_{2D}$, $\omega_R$) of 2D (a) and R (b) Raman bands in the tBLG sample (Fig. 1 in the main text) as a function of $n_{TG}$, which has good approximation to that in tBLG in low doping regime (Fig. 3). Empty blue (2D band) and solid pink (R band) circles are experimental data. These data are fitted with a phenomenological formula (see the discussion in the main text): $\omega = a + bn_{TG} + cn_{TG}^2 + dn_{TG}^3 + e|n_{TG}|^{3/2}$, where $\omega$ and $n_{TG}$ are in units of cm$^{-1}$ and $10^{13}$ cm$^{-2}$, respectively.[S8] The fitting parameters are summarized in Table S1.

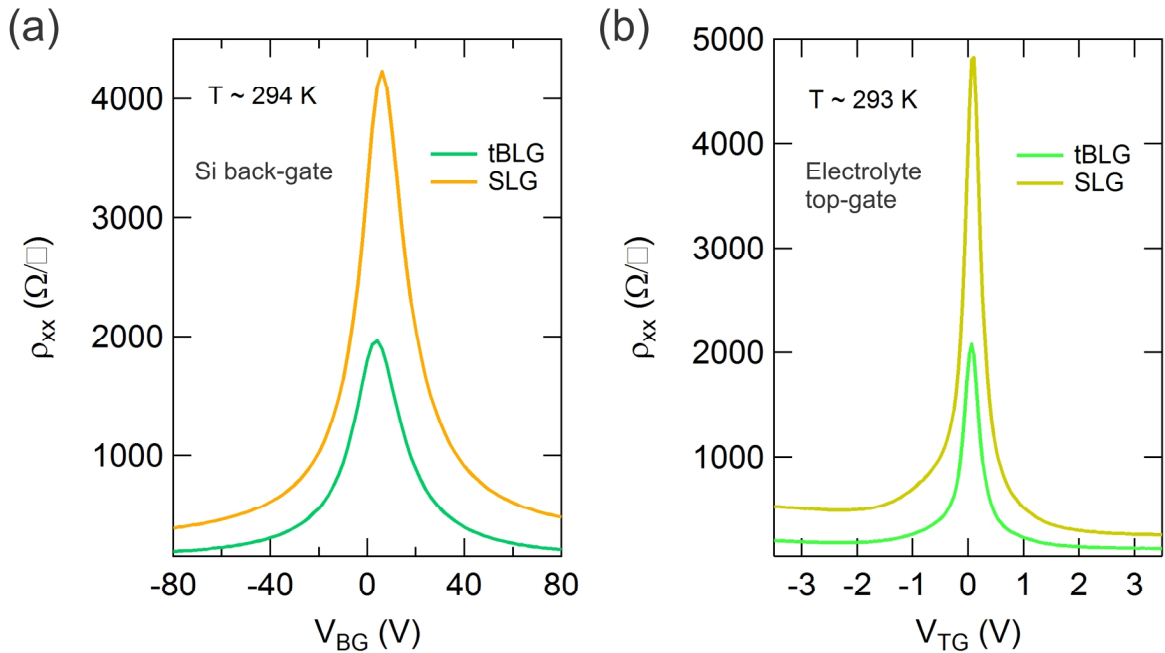

Figure S5. Field effect measurement (2D resistivity versus gate voltage) of the SLG and tBLG devices using (a) SiO$_2$/Si back-gate (before application of ion gel dielectrics) and (b) ion gel top-gate dielectrics. The CNP voltage (V$_D$) of the SLG and tBLG using the back-gate is about 6 V and 4 V (slightly p-doped), respectively. The V$_D$ of both devices is ~ 0 V when using the electrolyte top-gate. Hysteresis could cause a shift in V$_D$ on the order of ~ 0.4 V. The gate voltage sweep direction is from negative to positive.

|         | *a*    | *b*    | *c*    | *d*    | *e*    |
|---------|--------|--------|--------|--------|--------|
| **R band**  | 1492.1 | -1.708 | 1.297  | -0.059 | -2.245 |
| **2D band** | 2700.4 | -0.509 | -1.179 | -0.097 | 0.923  |

Table S1. Fitting parameters of R and 2D Raman peak frequencies for the tBLG based on a phenomenological formula (see the discussion in the main text): $\omega = a + bn_{TG} + cn_{TG}^2 + dn_{TG}^3 + e|n_{TG}|^{3/2}$, which describes the shift of phonon frequency $\Delta\omega = \omega - a$ as a function of carrier density $n_{TG}$ (in the units of $10^{13}$ cm$^{-2}$).[S8]